\documentclass[aps,prl,twocolumn,nopacs,superscriptaddress]{revtex4}
\usepackage{graphicx}
\usepackage{verbatim}
\usepackage{amsmath}
\usepackage{sidecap}
\usepackage{epstopdf}
\usepackage{braket}

\begin{document}

\title{Irreversible proliferation of magnetic moments at cleaved surfaces of the topological Kondo insulator SmB$_6$}
%\title{Irreversible thermally modulated evolution promoting magnetism at cleaved surfaces of topological Kondo insulator SmB$_6$}
%\title{Irreversible evolution promoting magnetism at cleaved surfaces of topological Kondo insulator SmB$_6$}
%\title{Symmetries, energetics and time dependence of the topological Kondo surface of SmB$_6$}
%\title{Irreversible thermal evolution of the topological surface Kondo lattice in SmB$_6$}
%\title{Irreversible thermal evolution of the topological surface state Kondo lattice in SmB$_6$}

\author{Haowei He}
%\thanks{These authors contributed equally to this work.}
\affiliation{Department of Physics, New York University, New York, New York 10003, USA}
\author{Lin Miao}
\affiliation{Advanced Light Source, Lawrence Berkeley National Laboratory, Berkeley, CA 94720, USA}
\affiliation{Department of Physics, New York University, New York, New York 10003, USA}
%\thanks{These authors contributed equally to this work.}
\author{Edwin Augustin}
\author{Janet Chiu}
\author{Surge Wexler}
\author{S. Alexander Breitweiser}
\affiliation{Department of Physics, New York University, New York, New York 10003, USA}
%\author{Yishuai Xu}
\author{Boyoun Kang} % (GIST) 
%(Boyoun Kang, bykang@gist.ac.kr)
\author{B. K. Cho} %  (GIST) 
%  chobk@gist.ac.kr
\affiliation{School of Materials Science and Engineering, Gwangju Institute of Science and Technology (GIST), Gwangju 61005, Korea}
\author{Chul-Hee Min}
% cmin@physik.uni-wuerzburg.de
\author{Friedrich Reinert}
\affiliation{Experimentelle Physik VII and R\"ontgen Research Center for Complex Materials (RCCM),
Universit\"at W\"urzburg, 97074 W\"urzburg, Germany}
\author{Yi-De Chuang}
\author{Jonathan Denlinger}
\affiliation{Advanced Light Source, Lawrence Berkeley National Laboratory, Berkeley, CA 94720, USA}
\author{L. Andrew Wray}
\email{lawray@nyu.edu}
\thanks{Corresponding author}
\affiliation{Department of Physics, New York University, New York, New York 10003, USA}

\begin{abstract}

The compound SmB$_6$ is the best established realization of a topological Kondo insulator, in which a topological insulator state is obtained through Kondo coherence. Recent studies have found evidence that the surface of SmB$_6$ hosts ferromagnetic domains, creating an intrinsic platform for unidirectional ballistic transport at the domain boundaries. Here, surface-sensitive X-ray absorption (XAS) and bulk-sensitive resonant inelastic X-ray scattering (RIXS) spectra are measured at the Sm N$_{4,5}$-edge, and used to evaluate electronic symmetries, excitations and temperature dependence near the surface of cleaved samples. The XAS data show that the density of large-moment atomic multiplet states on a cleaved surface grows irreversibly over time, to a degree that likely exceeds a related change that has recently been observed in the surface 4f orbital occupation.

%***Previous investigations have established a roughly temperature-independent time evolution at temperatures below T$\lesssim$140K. Here we find that the time evolution is very rapid at higher temperature, but evolves to the same point at low or high temperature. Exposure to air is found to cause further oxidation.  
%***The measurements suggest that Sm sites in the initial state are predominantly small-moment, in line with the expectation that the moments will be heavily screened***
%***also note that: although the correspondence of multiplet predictions with RIXS features is mediocre, the close correspondence between multiplet numerics and features in the XAS profile is very encouraging. Together, the RIXS and XAS data suggest that a more sophisticated model is needed for the low energy physics, however a simple atomic model may be sufficient to approximate the resonant matrix elements of the RIXS process.
%reduces the density of $J=0$ sites, increasing the density of spinors in the topological surface Kondo lattice.

\end{abstract}

% \pacs{73.20.At, 73.20.Hb, 72.10.Fk}

\date{\today}

\maketitle

%By this means, the site- (s) ??resolved Bloch spectral density N_s (??E; k).

The topological Kondo insulator (TKI) state is a variant of the topological insulator state \cite{pred1,ColemanReview,FuOriginal,ZahidMooreReview}, in which a topologically ordered insulating electronic band structure is obtained from Kondo physics. The realization of a TKI state in mixed-valent SmB$_6$ was strongly indicated by early theoretical investigations \cite{pred1,pred2}, and has now been established through direct measurement of the topological surface states via angle resolved photoemission \cite{MadhabARPES,FengARPES,ShiSpinTexture,Kondo110K,ReinertChargeFluctuations,JonathanSmB6PEreview} and transport studies \cite{TKItransport,TKIsurfaceFM}. Strong evidence has recently been found suggesting that the surface of polished SmB$_6$ samples can also host ferromagnetic domains \cite{TKIsurfaceFM}, a property that is theoretically associated with exotic axion electrodynamics, an inverse spin-galvanic effect, and ballistic one dimensional transport channels at domain boundaries \cite{magTI1,MagTI2,MagTI3,ZahidMooreReview}. Moreover, surface sensitive X-ray photoemission (XPS) measurements have shown that the surface 4f occupation evolves irreversibly towards 4f$^5$ as a function of time following cleavage in ultra high vacuum (UHV) \cite{surfaceValencePhMag}. Here, multiplet-dominated X-ray absorption spectroscopy (XAS) and resonant inelastic X-ray scattering (RIXS) measurements in the vacuum ultraviolet (VUV) regime are used as a symmetry-sensitive probe to map the Sm N$_{4,5}$-edge excitations and show that a similarly large change in the density of large-moment samarium sites accompanies this time evolution. This evolution is consistent with expectations for the transition from a Kondo insulating state to magnetism, and represents a means for incrementally tuning the strength of the surface magnetic instability.

\begin{figure}[t]
\includegraphics[width = 8cm]{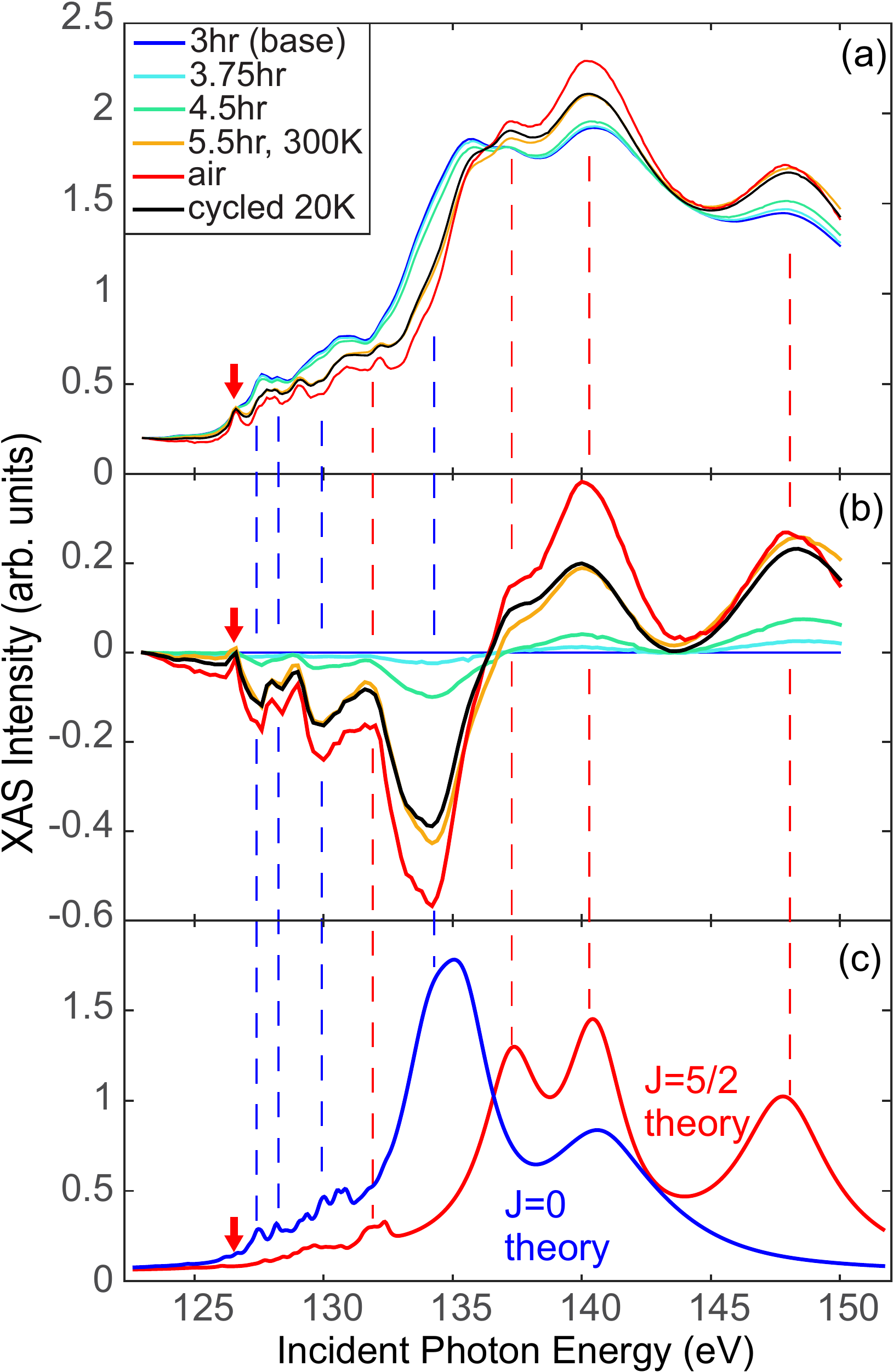}
\caption{{\bf{Irreversible proliferation of local moments}}. (a) X-ray absorption measurements of the surface of topological Kondo insulator SmB$_6$ show increasingly rapid irreversible surface changes as the sample is aged. The sample was maintained below 100K for the first 4.5 hr after cleavage, and was then heated to 300K. The spectrum that resulted after cycling back to low temperature is shown in black, and a sample that had been stored in ambient air is also shown (red curve). (b) Difference curves, subtracting the base (t=3 hr) XAS profile. (black curve) A sample that has been cycled to room temperature continues to resemble the high temperature spectrum upon cooling back to T=20K. The trend of changes from aging resembles (red curve) the result of oxidation from air exposure. (c) Numerical simulations of magnetically inert J=0 (4f$^6$) sites and large-moment J=5/2 (4f$^5$) sites. Red arrows highlight the anomalous h$\nu$$=$126.6 eV feature.}
\label{fig:ChangingKondoLattice}
%\vspace{-0pt}
\end{figure}

%\section{Methods}

% and Sm$_{0.98}$La$_{0.02}$B$_6$ 
Measurements were performed at the beamline 4.0.3 (MERLIN) RIXS endstation (MERIXS) \cite{YiDeDetector,YiDeSRN} at the Advanced Light Source (ALS), Lawrence Berkeley National Laboratory.  Large single crystals of SmB$_6$ were grown by the Al flux method as in Ref. \cite{surfaceValencePhMag}, cleaved at low temperature, and maintained at a UHV pressure of approximately 3$\times$10$^{-10}$ Torr. The photon beam had a grazing 30$^o$ or angle of incidence to the cleaved [001] sample face, and scattered photons were measured at 90$^o$ to the incident beam trajectory. XAS was measured using the total electron yield (TEY) method, and the expected penetration depth of measurements is roughly d$\lesssim$2 nm for XAS \cite{universalCurve} and d$\sim$10-30 nm for RIXS \cite{XrayPath}. To minimize sensitivity to surface inhomogeneity, the beam profile on the sample was configured as a very broad strip with dimensions of roughly $10 \times 600 \mu m^2$ (similar results from additional cleaves are shown in the online Supplimental Material (SM) \cite{SM}).

%Temperature stabilization on each heating step took approximately 20 minutes, and 

%lifetime [0.6, 135, 3, 139, 4]

Atomic multipet simulations were performed with typical renormalization values for the multipolar Slater-Condon interaction parameters \cite{SM}. Similar multiplet models that focus on Sm f-electrons, disregarding the itinerant 5d electron gas, have been remarkably successful in reproducing XAS and XPS features of SmB$_6$ \cite{multipletSmB6_PES2015,JonathanSmB6PEreview}. The multiplet ground states have f-electron angular momentum quantum numbers of (4f$^5$) J=5/2 and (4f$^6$) J=0, representing the presence or absence of a hole in the J=5/2 4f bands. The atomic multiplet picture is expected to be most accurate as a description of the resonance states, which are dominated by extremely strong angular momentum coupling between the 4d core hole and 5d electrons on the scattering site (a $\sim$20eV combined energy scale). Multiplet state energetics in the VUV are defined in terms of coherent local moment symmetries. In contrast to the previous study of 4f charge density at the SmB$_6$ surface \cite{surfaceValencePhMag}, atomic multiplet measurements in the VUV are sensitive to the \emph{coherent multi-particle symmetry} of electrons in hybridized electronic orbits involving both the scattering site and neighboring atoms \cite{WrayFrontiers,EdwinProc}. This multiplet symmetry can be thought of as the `nominal valence' state defining local moment degrees of freedom, and can deviate significantly from the  atomically resolved charge density.

%We will therefore refer to the 4f$^5$ and 4f$^6$ simulations as representing J=5/2 and J=0 scenarios respectively, to emphasize the role of these states with respect to a Kondo lattice picture in which just one 4f band hybridizes with itinerant 5d states at the Fermi level \cite{pred1,pred2}.

%\section{Results}

%Figure 1:

The samarium N$_{4,5}$-edge XAS spectrum of a pristine SmB$_6$ surface that has been recently cleaved and maintained at T$<$100K within UHV is shown in Fig. \ref{fig:ChangingKondoLattice}(a) (blue curve). To facilitate comparison of features, the non-resonant background has been aligned beneath the resonance at $h\nu$=123 eV, and the curves have been set to have the same integrated area. The spectrum contains a large number of features, and the higher energy features have broader line shapes, as is the typical trend for core hole lifetimes at a multiplet-split resonance \cite{HaverkortAutoIon,KotaniIdea,KotaniIdea2,WrayNiO,WrayFrontiers,WrayRIXSinterference,EdwinProc}. Relatively sharp line shapes are observed in the incident energy range from 126-133, suggesting that there may be a charge transfer threshold at h$\nu$$\sim$133 eV \cite{HaverkortAutoIon,EdwinProc}.

The spectrum changes pronouncedly as the sample is aged. Difference spectra in Fig. \ref{fig:ChangingKondoLattice}(b) show that as time progresses, spectral intensity shifts into higher energy features at $h\nu>136$ eV, and the sharper 126-133 eV features shift into a new spectral pattern that bears little resemblance to that seen initially. Even though the time interval between each pair of successive curves is approximately the same (45 to 60 min), the magnitude of the change is significantly larger for the interval from 4.5 to 5.5 hr, in which the sample was heated to room temperature. Cycling back to low temperature (black curve) resulted in only small quantitative changes. Later scans did not reveal continued changes, however the aging trend can be taken further still by exposing the sample surface to air (see red curve), suggesting that aging the surface is promoting changes in the f-electron count that resemble oxidation. The fractional change in feature intensities after thermal cycling is dramatic, and ranges from 10-50$\%$ throughout most of the spectrum.

To identify the physical significance of the surface evolution, a multiplet simulation in Fig. \ref{fig:ChangingKondoLattice}(c) shows the features expected in Sm N$_{4,5}$-edge XAS from J=0 and J=5/2 sites. Dashed drop-lines highlight an excellent qualitative match between the regions that lose intensity during the aging process and the J=0 features, while the J=5/2 features correlate with a gain (or reduced loss) of intensity. Only one feature is clearly anomalous in this analysis. The lowest energy XAS peak at h$\nu$$=$126.6 eV is not reproduced by either the J=0 or the J=5/2 multiplet calculation, and has temperature dependence consistent with a J=5/2 symmetry attribution.

%Though the correspondence with theory is not quantitative enough to provide a precise estimate of the numerical ratio of J=5/2 and J=0 sites, large thermal changes in the XAS features below 135eV are suggestive of a change of at least 30$\%$ from T=20 to 300K.  More moderate changes at higher energy can be approximated with a change of $\sim$15$\%$. The higher energy range is generally subject to greater modeling error, and we conclude that the change is likely to be at least $\gtrsim$20$\%$.

%The trend towards fewer 4f-electrons with increasing temperature is consistent with the trend in bulk valence (cite many***including Jonathan), however the irreversible nature of the trend observed here i

\begin{figure}[t]
\includegraphics[width = 8cm]{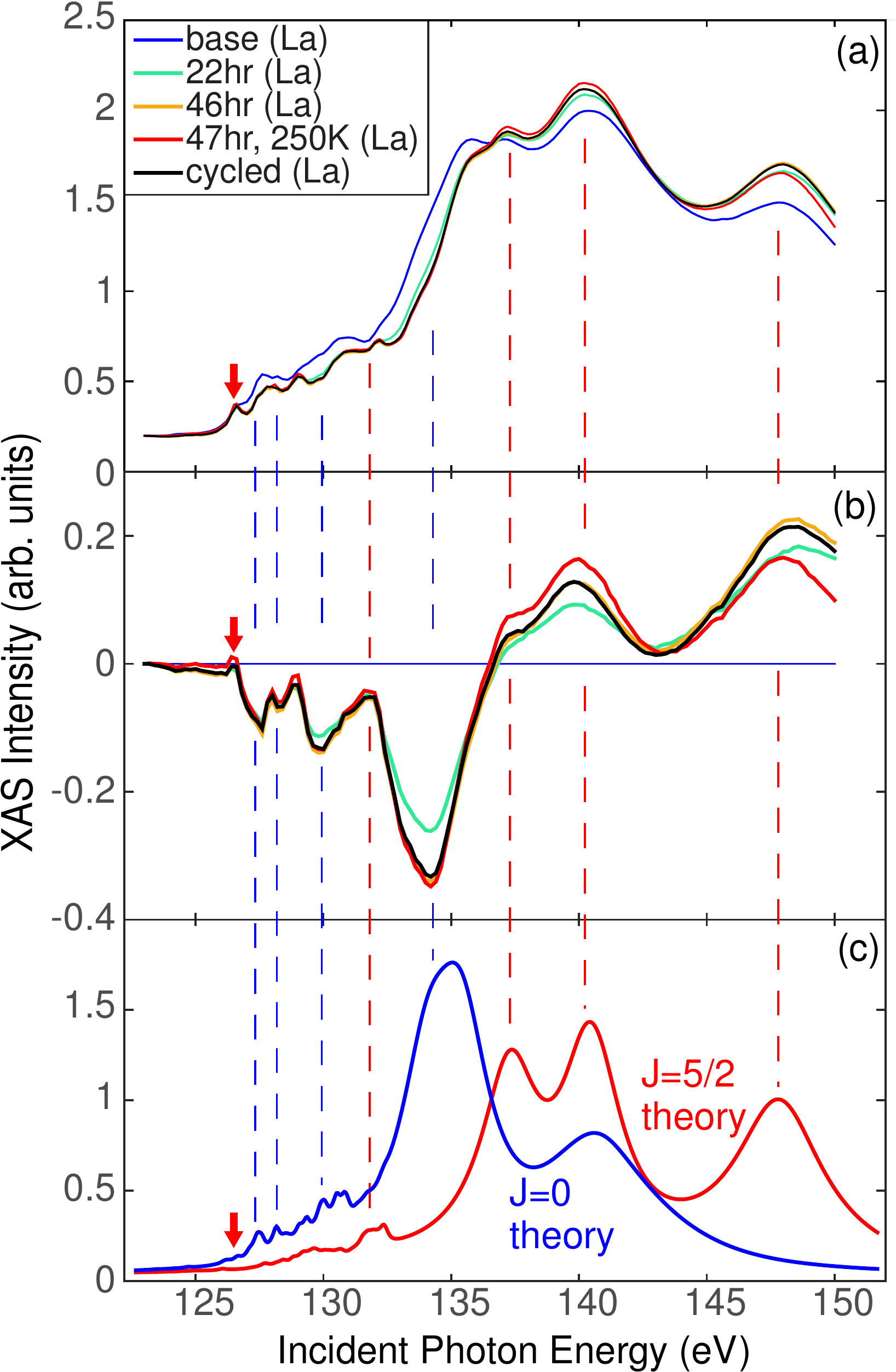}
\caption{{\bf{Slow surface evolution at low temperature}}. (a) X-ray absorption measurements as a function of time for slightly doped Sm$_{0.98}$La$_{0.02}$B$_6$ maintained beneath T=100K. High temperature (T=200K) and cycled T$<$100K post-evolution curves are also shown. The final cycled curve (black) is nearly identical to the 46 hr curve (yellow), and the two cannot be easily distinguished by eye. (b) Difference curves, subtracting the base (t$\sim 3$ hr) XAS profile. (c) Numerical simulations of magnetically inert J=0 (4f$^6$) sites and large moment J=5/2 (4f$^5$) sites. Red arrows highlight the anomalous h$\nu$$=$126.6 eV feature.}
\label{fig:ChangingKondoLatticeLa}
%\vspace{-0pt}
\end{figure}

\begin{figure*}[t]
\includegraphics[width = 17.8cm]{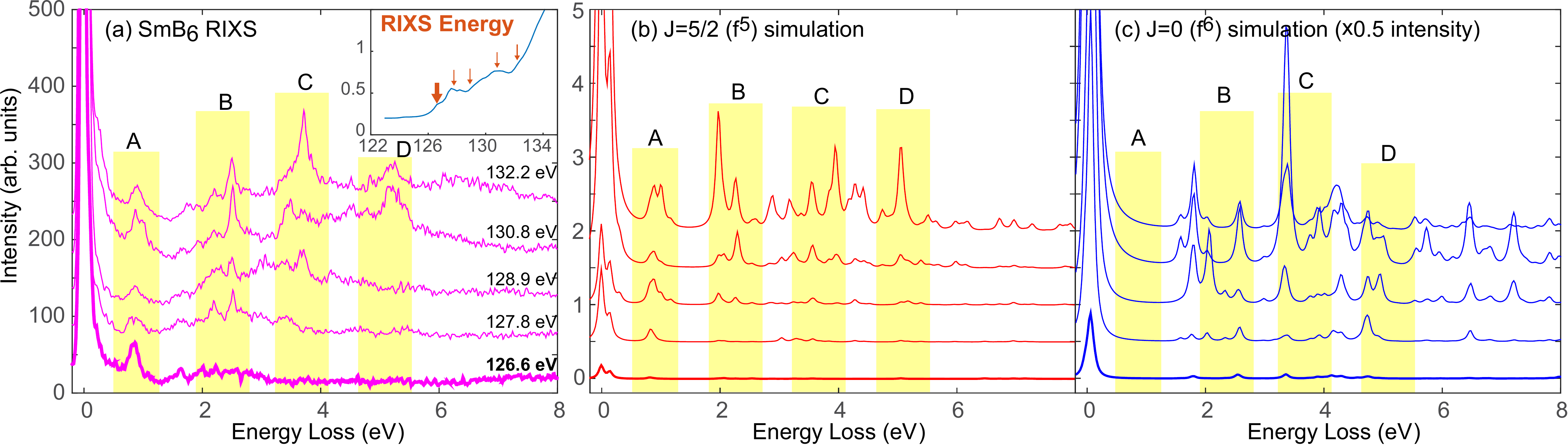}
\caption{{\bf{Sharp inelastic excitations}}. (a) RIXS spectra of recently cleaved SmB$_6$ maintained at T=20K using the labeled incident photon energies, which are also indicated on (inset) the T=20K XAS profile. Numerical simulations are presented for RIXS from (b) J=5/2 (4f$^5$) and (c) J=0 (4f$^6$) local moment sites. Curves at the anomalous h$\nu$$\sim$126.6 eV resonance peak are plotted with a thicker line, and four energy loss regions (A-D) that contain strong scattering within the J=5/2 simulation are highlighted in yellow.}
\label{fig:RIXSanalysis}
%\vspace{-0pt}
\end{figure*}

Similar XAS measurements have also been performed as a function of time on an electron doped sample with the composition Sm$_{0.98}$La$_{0.02}$B$_6$ (see Fig. \ref{fig:ChangingKondoLatticeLa}). In this case, the sample was maintained at T$<$100K for a much longer 46 hr period, and the measurements confirm earlier observations that the aging process at low temperature proceeds on a time scale longer than 1 day \cite{surfaceValencePhMag}. Heating to T=250K produced small changes that were fully reversed upon cycling back to T$<$100K. This is in contrast to the more rapid measurement on undoped SmB$_6$, and suggests that the physical end-point of low temperature aging is the same as the rapidly achieved end-point of room temperature aging.

Lanthanum is expected to act as a net electron donor, entering an ionization state much closer to 3+ as compared to Sm. Doping into the Sm 4f orbitals of roughly 0.3e$^-$/La atom is attributed from susceptibility studies \cite{tempAndValence}. The initial Sm$_{0.98}$La$_{0.02}$B$_6$ XAS spectrum is nearly identical to the slightly aged t=4.5 hr spectrum of undoped SmB$_6$, and the final aged (cycled) curve is qualitatively identical to XAS from the fully aged undoped SmB$_6$ sample. The fact that J=5/2 features in the base (t$\sim 3$ hr) spectrum of Sm$_{0.98}$La$_{0.02}$B$_6$ are more prominent than in the low temperature spectrum of SmB$_6$ is at odds with the identification of La as an electron donor, and suggests some variability in the nature of the cleaved surface (this is confirmed in Fig. S3 of the SM \cite{SM}).

\begin{figure}[t]
\includegraphics[width = 8cm]{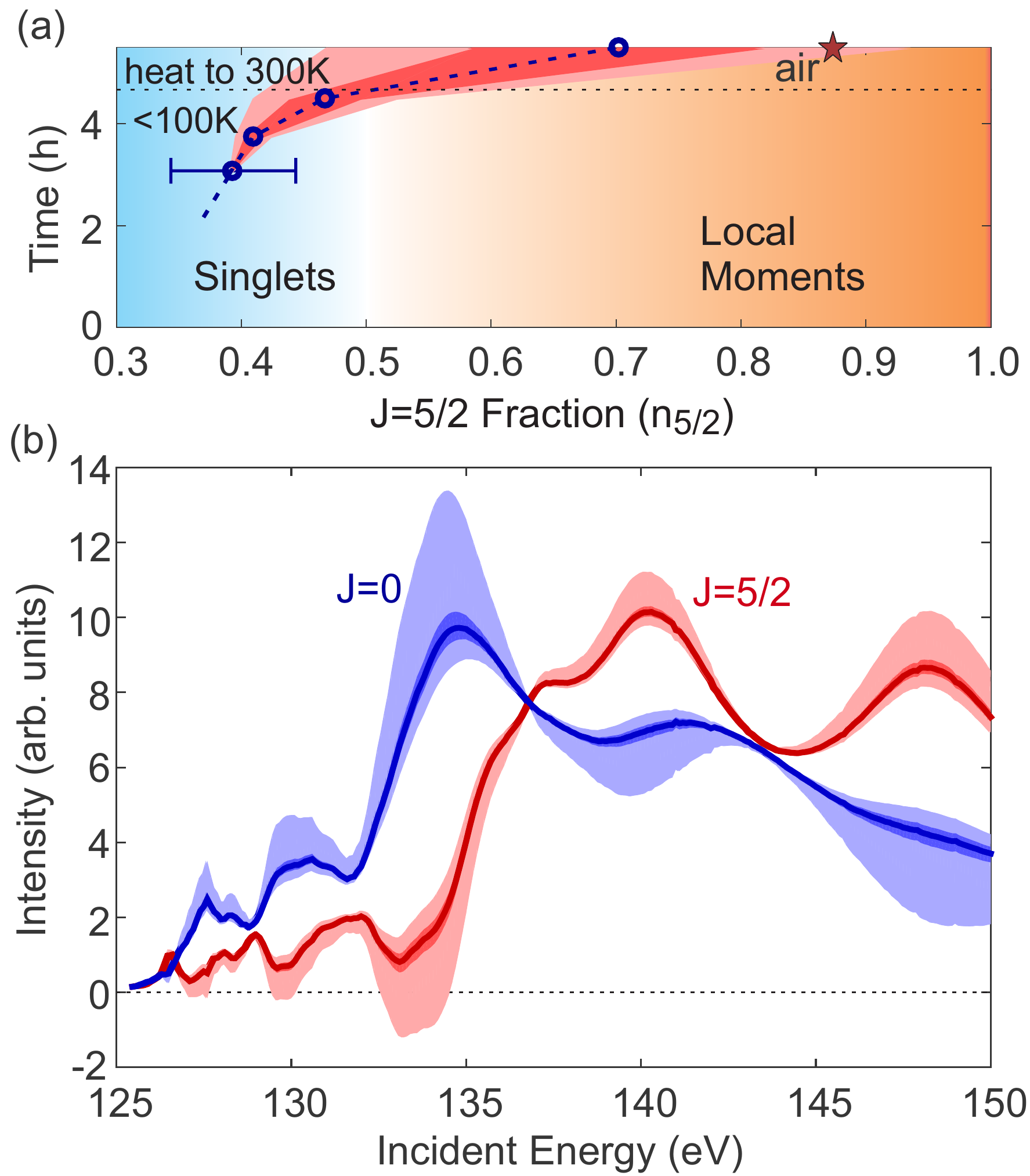}
\caption{{\bf{A path to magnetism}}. (a) The surface aging trend of the J=5/2 site fraction is estimated from extrapolations of the pure J=5/2 and J=0 single-site XAS curves \cite{SM}. The sample was heated to 300K after 4.5 hours, resulting in more rapid aging. The estimated density of J=5/2 sites in an air-exposed sample is indicated with a star. (b) Algebraically determined pure J=5/2 and J=0 XAS curves. Shaded regions indicate variability based on (darker) $\pm0.1$ and (lighter) $\pm0.4$ differences in the estimated ratios of monovalent site densities at t=3 hr and 5.5 hr (i.e. $\frac{n_{5/2}(5.5 hr)}{n_{5/2}(3 hr)}$). Resulting error in the aging trend curve is indicated by the same shading in panel (a).}
\label{fig:toMagnetism}
%\vspace{-0pt}
\end{figure}

Measuring resonant inelastic scattering at the photon energies used for XAS reveals a wide range of excitations, some of which are quite sharp in energy, as seen in Fig. \ref{fig:RIXSanalysis}(a). Prominent features at E=0.9, 2.5, 3.7 and 5.2 eV appear to be Raman-like (non-dispersive), however the upper bound of intensity on the energy loss axis disperses with a slope similar to 1 as a function of incident energy, starting from $\sim$3 eV at an incident energy of h$\nu$=126.6 eV, as expected for scattering scenarios that involve interplay with degrees of freedom that have similar energetics with or without a core hole present. This suggests that a truly accurate model of the RIXS excitations must incorporate significant non-local physics, going beyond the degrees of freedom on a single scattering site. The RIXS spectrum and optical conductivity \cite{allenOptical1977,RussianOC,newOC} both picks up intensity from roughly 1.5eV, suggesting the onset of a large density of itinerant continuum states, however the optical data contain no candidates for \emph{any} of the energetically sharp features seen by RIXS.

In spite of not fulfilling this requirement, the atomic multiplet model is still expected to be accurate for certain relatively localized excitations, and to yield accurate matrix elements of the `direct RIXS' scattering process \cite{AmentRIXSReview,WrayRIXSinterference}. One can attribute a tentative correspondence between energy loss regions with large intensity in the J=5/2 simulation and the intensity seen by RIXS (see highlighted regions in Fig. \ref{fig:RIXSanalysis}(a-b)). The feature at E$\sim$0.9 eV gives a particularly close correspondence, and has no competing interpretation within the J=0 simulation (Fig. \ref{fig:RIXSanalysis}(c)). This E$\sim$0.9 eV mode is actually a collection of closely spaced features, with additional peaks visible in the $h\nu=$127.8 and 130.8 eV curves. Lower energy excitations visible in the simulations are not resolved from the elastic line, which is strong due to the broad off-angle tail of specular reflection in the VUV \cite{WrayFrontiers}. The lack of easily identifiable J=0 derived features may indicate that single-atom excitations on 4f$^6$ sites are shorter lived (i.e. broader), possibly because they can easily delocalize into 4f$^5$\textbf{X} states, where `\textbf{X}' indicates an electron that has entered a more delocalized band symmetry. The anomalous $h\nu=126.6$ eV resonance, which does not occur in our multiplet simulations, resonates primarily with the 0.9 eV feature, corroborating identification of the $h\nu=126.6$ eV resonance with scattering from a J=5/2 site.

%Note that it is unsurprising to see large deviations in RIXS, because the model is not very accurate for low energy states, as discussed in the Methods Section.

%\section{Discussion}

A sufficiently large increase in the density of J=5/2 sites in the mixed-valent samarium lattice is expected to destabilize the Kondo insulating state and induce magnetism, as has been seen in high pressure studies \cite{pressureNf,pressure2,pressureNickButch}. The amplitude of the surface change can be evaluated from the XAS data in Fig. \ref{fig:ChangingKondoLattice}, if we adopt the approximation that the spectra can be broken down into linear combinations of pure J=5/2 and J=0 curves (see derivations in the SM \cite{SM}). The accuracy of this approximation is supported in the present case by the observation of a fairly stable isospectral (constant intensity) point at 136.7-137eV for all warming curves.

Monovalent J=5/2 and J=0 XAS spectra algebraically obtained from comparing SmB$_6$ XAS measurements at the pristine (t=3 hr) and aged (cycled) surface are plotted in Fig. \ref{fig:toMagnetism}(b). Shaded regions represent error margins as described in the caption, and are associated with different extrapolated changes in the population of J=5/2 and J=0 sites over the aging period. The outer shaded boundaries represent extreme scenarios that lead to clear anomalies in the extrapolated monovalent curves, such as duplicated features and negative XAS intensity \cite{SM}. The best fit J=0 XAS curve is obtained with the assumption that the J=5/2 population grows by a remarkable 60-80$\%$. Growth estimates beneath 40$\%$ lead to multiple significant artifacts in the extrapolated XAS spectrum. Based on the extrapolated XAS curves, we can assess that a single-atom multiplet calculation provides many matrix elements of the N$_{4,5}$-edge resonance process, but is of marginal use for understanding the RIXS excitations, and omits XAS features that appear to be associated with coupling to itinerant states.

High pressure studies have found that a new magnetic ground state \cite{pressureNf,pressure2,pressureNickButch} is realized in bulk SmB$_6$ beyond a crossover point thought to occur when the fractional density of 4f$^5$ sites reaches roughly $n_{5/2}\sim0.65$ \cite{pressureNf,pressureNickButch}. Combined examination of the extrapolated J=5/2 and J=0 curves enables an algebraic derivation of the t=3 hr J=5/2 site density as $n_{5/2}$=0.39$\pm$0.05 \cite{SM}, with a surface evolution plotted in Fig. \ref{fig:toMagnetism}(a). In distinguishing $n_{5/2}$ from the nominal bulk 4f occupancy, which is thought to be $n_f\lesssim 5.5$ \cite{pressureNf,pressureNickButch,JonathanSmB6PEreview,XASvalenceTdep,JonathanARPES_DMFT}, this picture suggests that the transition is one from a singlet-dominated low temperature regime, with fewer observed large-moment sites than the 4f occupancy alone would suggest, to an aged surface in which local moments are less screened. This scenario matches the attributed behavior as the crystal is driven into a magnetic state under pressure \cite{pressureNf,pressureNickButch}. We note that the nature of the magnetic order achieved under pressure is not definitively understood, however a proximate ferromagnetic state can be achieved from 1$\%$ Fe doping at ambient pressure \cite{Fe1pctMag}. The air-exposed end point of $n_{5/2}\sim0.9$ is consistent with the proposed interpretation of an earlier soft X-ray M-edge XAS measurement \cite{CdopingAndXAS}.

%The earlier interpretation is difficult to compare closely with, as it represents a very different regime for the interplay of surface effects (such as known reconstructions \cite{STM1,flashed}) with multiplet energetics, and the paper does not establish correspondence with theoretically predicted atomic multiplet spectral structures. % and the intrinsic difference between TEY and total fluorescence yield (TFY) XAS matrix elements.

%%%%%%%%%% revised to here %%%%%%%%%%%%%%

%The recent systematic measurement of surface 4f occupancy in Ref. \cite{surfaceValencePhMag} reported a smaller effect (50\%$ increase in 4f$^5$ site density) and slower time evolution than is reported here, and did not show as clear an effect of temperature on the rate of surface aging. The intrinsic difference between 

%coherent multiplet state of electrons on the scattering site and neighboring atoms

% greater surface sensitivity and a larger measured temperature range in the current study, Surface sensitivity in the present work is determined from the escape depth of $E<150$ eV electrons as compared to $\sim$400 eV electrons in Ref. \cite{surfaceValencePhMag}.

A comprehensive interpretation of the mechanism behind the observed irreversible surface changes is beyond the scope of the present study. Such a theory may need to address multiple factors including the correlated electronic structure, the intrinsically polar and anisotropic nature of cleaved surfaces, and the presence of multiple known surface reconstructions \cite{JonathanSmB6PEreview,STM1,flashed}. The bulk of the sample also evolves towards a lower 4f occupancy with increasing temperature \cite{pressureNf,JonathanSmB6PEreview,XASvalenceTdep,JonathanARPES_DMFT,Kondo110K}. However, the bulk-sensitive RIXS spectrum undergoes no easily visible changes as a function of temperature \cite{SM}, suggesting that surface aging is a far more dramatic effect.  The La doped sample is thought to have roughly identical bulk temperature dependence \cite{tempAndValence}.

%These irreversible changes in surface local moments are likely related to the known [2x1] reconstruction as well as other more complex reconstructions, which cover the SmB$_6$ surface upon cleaving in ultra high vacuum (UHV) at low temperature \cite{STM1}. The [2x1] reconstruction has also been documented at the surface of samples that have been flashed to T=1400C \cite{flashed}.

%***also note that: although the correspondence of multiplet predictions with RIXS features is mediocre, the close correspondence between multiplet numerics and features in the XAS profile is very encouraging. Together, the RIXS and XAS data suggest that a more sophisticated model is needed for the low energy physics, however a simple atomic model may be sufficient to approximate the resonant matrix elements of the RIXS process.

These resonant X-ray absorption results show that the density of large-moment Sm sites in the top $\lesssim$2 nm of cleaved SmB$_6$ more than doubles as the surface ages in UHV. The increase can be accelerated by heating to room temperature, and is taken further through exposure to air. Lower energy excitations of SmB$_6$ are mapped with RIXS, providing a window into dynamics and energetics of the valence electrons, which will serve as a reference for future theoretical and spectroscopic investigations. The large surface changes seen by electron yield XAS are distinct from the lack of extraordinary temperature dependence at depths of 10-30 nm, as evaluated from RIXS spectra. The increased density of large-moment sites on aged samples provides a plausible explanation for the recent observation of ferromagnetic domains at a polished SmB$_6$ surface \cite{TKIsurfaceFM}. More generally, the apparent sensitivity of the surface evolution to temperature and the gas environment provides mechanisms for control of the surface properties, to achieve a desired interplay between surface magnetic moments and the topologically ordered bulk electronic structure.

%and suggests a thermally tuned mechanism for UHV studies to induce - or temporarily avoid - the magnetic state.

%, but is irreversibly pinned by the [2$\times$1] surface reconstruction or some other structural change.

\textbf{Acknowledgements:} The Advanced Light Source is supported by the Director, Office of Science, Office of Basic Energy Sciences, of the U.S. Department of Energy under Contract No. DE-AC02-05CH11231. C.H.M. was supported by the DFG (through SFB 1170 ``ToCoTronics", projects C06). We are grateful for productive discussions with H. Dehghani and J. Hoffman.


\begin{thebibliography}[

%\bibitem{residualFS} B. S. Tan \emph{et al.}, Unconventional Fermi surface in an insulating state, Science \textbf{349}, 287 (2015).
%?the system is thought to be on the border of quantum criticality?

\bibitem{FuOriginal} L. Fu, C. L. Kane, and E. J. Mele, Topological Insulators in Three Dimensions, Phys. Rev. Lett. \textbf{98}, 106803 (2007).
\bibitem{ZahidMooreReview} M. Z. Hasan and C. L. Kane, Colloquium: Topological insulators, Rev. Mod. Phys. \textbf{82}, 3045 (2010).
\bibitem{ColemanReview} M. Dzero, J. Xia, V. Galitski, and P. Coleman, Topological Kondo Insulators, Annu. Rev. Condens. Matter Phys. \textbf{7}, 249-280 (2016).
\bibitem{pred1} M. Dzero, K. Sun, V. Galitski, and P. Coleman, Topological Kondo Insulators, Phys. Rev. Lett. \textbf{104}, 106408 (2010).
\bibitem{pred2} Feng Lu, JianZhou Zhao, Hongming Weng, Zhong Fang, and Xi Dai, Correlated Topological Insulators with Mixed Valence, Phys. Rev. Lett. \textbf{110}, 096401 (2013).

\bibitem{MadhabARPES} M. Neupane, N. Alidoust, S-Y. Xu, T. Kondo, Y. Ishida, D. J. Kim, Chang Liu, I. Belopolski, Y. J. Jo, T-R. Chang, H-T. Jeng, T. Durakiewicz, L. Balicas, H. Lin, A. Bansil, S. Shin, Z. Fisk, and M. Z. Hasan, Surface electronic structure of the topological Kondo-insulator candidate correlated electron system SmB$_6$, Nat. Comm. \textbf{4}, 2991 (2013).
\bibitem{FengARPES} J. Jiang, S. Li, T. Zhang, Z. Sun, F. Chen, Z.R. Ye, M. Xu, Q.Q. Ge, S.Y. Tan, X.H. Niu, M. Xia, B.P. Xie, Y.F. Li, X.H. Chen, H.H. Wen, and D.L. Feng, Observation of possible topological in-gap surface states in the Kondo insulator SmB$_6$ by photoemission, Nat. Comm. \textbf{4}, 3010 (2013).
\bibitem{ShiSpinTexture} N. Xu, P. K. Biswas, J. H. Dil, R. S. Dhaka, G. Landolt, S. Muff, C. E. Matt, X. Shi, N. C. Plumb, M. Radovi\'c, E. Pomjakushina, K. Conder, A. Amato, S. V. Borisenko, R. Yu, H.-M. Weng, Z. Fang, X. Dai, J. Mesot, H. Ding, and M. Shi, Direct observation of the spin texture in SmB$_6$ as evidence of the topological Kondo insulator, Nat. Comm \textbf{5}, 4566 (2014).

\bibitem{Kondo110K} N. Xu, C. E. Matt, E. Pomjakushina, X. Shi, R. S. Dhaka, N. C. Plumb, M. Radovi\'c, P. K. Biswas, D. Evtushinsky, V. Zabolotnyy, J. H. Dil, K. Conder, J. Mesot, H. Ding, and M. Shi, Exotic Kondo crossover in a wide temperature region in the topological Kondo insulator SmB$_6$ revealed by high-resolution ARPES, Phys. Rev. B \textbf{90}, 085148 (2014).
\bibitem{ReinertChargeFluctuations} Chul-Hee Min, P. Lutz, S. Fiedler, B.?Y. Kang, B.?K. Cho, H.-D. Kim, H. Bentmann, and F. Reinert, Importance of Charge Fluctuations for the Topological Phase in SmB$_6$, Phys. Rev. Lett. \textbf{112}, 226402 (2014).
\bibitem{JonathanSmB6PEreview} J. D. Denlinger, J. W. Allen, J.-S. Kang, K. Sun, B.-I. Min, D.-J. Kim and Z. Fisk, SmB$_6$ Photoemission: Past and Present, JPS Conf. Proc. \textbf{3}, 017038 (2014).
\bibitem{TKItransport} D. J. Kim, J. Xia, and Z. Fisk, Topological surface state in the Kondo insulator samarium hexaboride, Nature Materials \textbf{13}, 466 (2014).
\bibitem{TKIsurfaceFM} Yasuyuki Nakajima, Paul Syers, Xiangfeng Wang, Renxiong Wang and Johnpierre Paglione, One-dimensional edge state transport in a topological Kondo insulator, Nature Physics \textbf{12}, 213 (2016).

\bibitem{magTI1} A. M. Essin, J. E. Moore, and D. Vanderbilt, Magnetoelectric polarizability and axion electrodynamics in crystalline insulators, Phys. Rev. Lett. \textbf{102}, 146805 (2009).
\bibitem{MagTI2} R. Li, J. Wang, X-L. Qi, and S.-C. Zhang, Dynamical axion field in topological magnetic insulators, Nature Physics \textbf{6}, 284-288 (2010).
\bibitem{MagTI3} I. Garate and M. Franz, Inverse spin-galvanic effect in the interface between a topological insulator and a ferromagnet, Phys. Rev. Lett. \textbf{104}, 146802 (2010).
\bibitem{surfaceValencePhMag} P. Lutz, M. Thees, T. R. F. Peixot, B. Y. Kang, B. K. Cho, Chul Hee Min and F. Reinert, Valence characterisation of the subsurface region in SmB$_6$, Phil. Mag. \textbf{96}, 3307-3321 (2016). %doi:10.1080/14786435.2016.1192724
%\bibitem{DamascelliPolarity} Z.-H. Zhu, A. Nicolaou, G. Levy, N. P. Butch, P. Syers, X. F. Wang, J. Paglione, G. A. Sawatzky, I. S. Elfimov, and A. Damascelli, Polarity-Driven Surface Metallicity in SmB$_6$, Phys. Rev. Lett. \textbf{111}, 216402 (2013).

%\bibitem{MagTI4} Yu, R. et al. Quantized anomalous Hall effect in magnetic topological insulators. Science 329, 61?64 (2010).
\bibitem{YiDeDetector} Y.-D. Chuang, J. Pepper, W. McKinney, Z. Hussain, E. Gullikson, P. Batson, D. Qian, M. Z. Hasan, High-resolution soft X-ray emission spectrograph at advanced light source, J. Phys. Chem. Solid \textbf{66}, 2173 (2005).
\bibitem{YiDeSRN} Y.-D. Chuang, L. A. Wray, J. Denlinger, and Z. Hussain, Resonant Inelastic X-ray Scattering Spectroscopy at MERLIN Beamline at the Advanced Light Source, Synchrotron Radiation News \textbf{25}, 23 (2012).

\bibitem{universalCurve} M. P. Seah, W. A. Dench, Quantitative electron spectroscopy of surfaces: A standard data base for electron inelastic mean free paths in solids. Surf. Interface. Anal. \textbf{1}, 2 (1979).
\bibitem{XrayPath} B. L. Henke, E. M. Gullikson, J. C. Davis, Atomic Data and Nuclear Data Tables \textbf{54}, 181 (1993).
\bibitem{SM} The online Supplemental Material includes technical notes pertaining to the measurements and the derivation of curves in Fig. 4.
\bibitem{multipletSmB6_PES2015} A. B. Shick, L. Havela, A. I. Lichtenstein, and M. I. Katsnelson, Racah materials: role of atomic multiplets in intermediate valence systems. Sci. Rep. \textbf{5}, 15429 (2015). %; doi: 10.1038/srep15429
 \bibitem{WrayFrontiers} L. A. Wray, S.-W. Huang, I. Jarrige, K. Ikeuchi, K. Ishii, J. Li, Z. Q. Qiu, Z. Hussain, and Y.-D. Chuang, Extending resonant inelastic X-ray scattering to the extreme ultraviolet, Frontiers in Physics \textbf{3}, 32 (2015).
\bibitem{EdwinProc} Edwin Augustin, Haowei He, Lin Miao, Yi-De Chuang, Zahid Hussain, and L. Andrew Wray, Charge transfer excitations in VUV and soft X-ray resonant scattering spectroscopies, DOI: 10.1016/j.elspec.2016.12.004 (2016).

\bibitem{HaverkortAutoIon} S. S. Gupta, J. A. Bradley, M. W. Haverkort, G. T. Seidler, A. Tanaka, and G. A. Sawatzky, Coexistence of bound and virtual-bound states in shallow-core to valence x-ray spectroscopies, Phys. Rev. B \textbf{84}, 075134 (2011).
\bibitem{WrayRIXSinterference} L. A. Wray, S.-W. Huang, Y. Xia, M. Z. Hasan, C. Mathy, H. Eisaki, Z. Hussain, and Y.-D. Chuang, Experimental signatures of phase interference and subfemtosecond time dynamics on the incident energy axis of resonant inelastic x-ray scattering, Phys. Rev. B \textbf{91}, 035131 (2015).
\bibitem{KotaniIdea} K. Okada, A. Kotani, H. Ogasawara, Y. Seino, and B. T. Thole, Auger decay of quasiparticle states: Calculation of the Ni 3p photoemission spectrum in NiCl$_2$, Phys. Rev. B \textbf{47} , 6203 (1993).
\bibitem{KotaniIdea2} H. Ogasawara, A. Kotani, and B. T. Thole, Lifetime effect on the multiplet structure of 4d x-ray-photoemission spectra in heavy rare-earth elements, Phys. Rev. B \textbf{50}, 12332 (1994).
\bibitem{WrayNiO} L. A. Wray, W. Yang, H. Eisaki, Z. Hussain, and Y.-D. Chuang, Multiplet resonance lifetimes in resonant inelastic x-ray scattering involving shallow core levels, Phys. Rev. B \textbf{86}, 195130 (2012).

\bibitem{tempAndValence} S. Gab\'ani, K. Flachbart, V. Pavl\'ik, T. Herrmannsd\"orfer, E. Konovalova, Y. Paderno, J. Brian\v{c}in, J. Trp\v{c}evsk\'a, Magnetic properties of SmB$_6$ and Sm$_{1-x}$La$_x$B$_6$ solid solutions, Czechoslovak Journal of Physics \textbf{52}, A225 (2002).

\bibitem{allenOptical1977} Allen, J. W., Valence Instabilities and Related Narrow-Band Phenomena, New York: Plenum Press, 1977. DOI: 10.1007/978-1-4615-8816-0$\_$64
\bibitem{RussianOC} A. I. Shelykh, K. K. Sidorin, M. G. Karin, V. N. Bobrikov, M. ,M. Korsljkova, V. N. Gurin, and I. A. Smirnov, Optical constants and electronic structure of LaB$_6$, EuB$_6$, SmB$_6$ single crystals prepared by the solution method, J. Less-Common Metals \textbf{82}, 291-296 (1981).
\bibitem{newOC} A. Tytarenko, K. Nakatsukasa, Y. K. Huang, S. Johnston, and E. van Heumen, From bad metal to Kondo insulator: temperature evolution of the optical properties of SmB$_6$, New J. Phys. \textbf{18}, 123003 (2016).


%%%%%%%%%%%%%%%%%%%%

\bibitem{AmentRIXSReview} L. J. P Ament, M. van Veenendaal, T. P. Devereaux, J. P. Hill, and J. van den Brink, Resonant inelastic x-ray scattering studies of elementary excitations, Rev. Mod. Phys. \textbf{83}, 705 (2011).
\bibitem{pressure2} A. Barla, J. Derr, J. P. Sanchez, B. Salce, G. Lapertot, B. P. Doyle, R. Rüffer, R. Lengsdorf, M. M. Abd-Elmeguid, and J. Flouquet, High-Pressure Ground State of SmB$_6$: Electronic Conduction and Long Range Magnetic Order, Phys. Rev. Lett. \textbf{94}, 166401 (2005).
\bibitem{pressureNf} J Derr, G Knebel, G Lapertot, B Salce, M-A M\'easson and J Flouquet, Valence and magnetic ordering in intermediate valence compounds: TmSe versus SmB$_6$, J. Phys: Condens. Matter \textbf{18} 2089-2106 (2006).
\bibitem{pressureNickButch} Nicholas P. Butch, Johnpierre Paglione, Paul Chow, Yuming Xiao, Chris A. Marianetti, Corwin H. Booth, and Jason R. Jeffries, Pressure-Resistant Intermediate Valence in the Kondo Insulator SmB$_6$, Phys. Rev. Lett. \textbf{116}, 156401 (2016).

\bibitem{XASvalenceTdep} Masaichiro Mizumaki, S. Tsutsui, and F. Iga, Temperature dependence of Sm valence in SmB$_6$ studied by X-ray absorption spectroscopy, J. Phys.: Conf. Series \textbf{176} 012034 (2009).

\bibitem{JonathanARPES_DMFT} J. D. Denlinger, J. W. Allen, J.-S. Kang, K. Sun, J.-W. Kim, J.H. Shim, B. I. Min, Dae-Jeong Kim, and Z. Fisk, Temperature Dependence of Linked Gap and Surface State Evolution in the Mixed Valent Topological Insulator SmB$_6$, Preprint available at ``https://arxiv.org/abs/1312.6637"

%\bibitem{pressure3} J. Derr \emph{et al.}, From unconventional insulating behavior towards conventional magnetism in the intermediate-valence compound SmB$_6$, Phys. Rev. B \textbf{77}, 193107 (2008)
\bibitem{Fe1pctMag} T. S. Altshuler and Yu. V. Goryunov, M. S. Bresler, Ferromagnetic ordering of iron impurities in the mixed-valence semiconductor SmB$_6$, Phys. Rev. B, \textbf{73}, 235210 (2006).

\bibitem{CdopingAndXAS} W. A. Phelan, S. M. Koohpayeh, P. Cottingham, J. W. Freeland, J. C. Leiner, C. L. Broholm, and T. M. McQueen, Correlation between Bulk Thermodynamic Measurements and the Low-Temperature-Resistance Plateau in SmB$_6$, Phys. Rev. X \textbf{4}, 031012 (2014).

\bibitem{STM1} S. R\"{o}{\ss}ler, T.-H. Jang, D.-J. Kim, L. H. Tjeng, Z. Fisk, F. Steglich, and S. Wirth, Hybridization gap and Fano resonance in SmB$_6$, Proc. Nat. Acad. Sci. \textbf{111}, 4798-4802 (2014).
%Tae-Hwan Janga, Dae-Jeong Kimb, L. H. Tjenga, Zachary Fiskb,1, Frank Steglicha, and Steffen Wirth
\bibitem{flashed} H. Miyazaki, T. Hajiri, T. Ito, S. Kunii, S. I. Kimura, Momentum-dependent hybridization gap and dispersive in-gap state of the Kondo semiconductor SmB$_6$, Phys. Rev. B \textbf{86}, 075105 (2012).

% James W. Allen2, Jeong-Soo Kang3, Kai Sun2, Byung-II Min4, Dae-Jeong Kim5, and Zachary Fisk


%\cite{HaverkortAutoIon}. However, other effects also contribute to the phenomenon of higher energy features having broader line shapes in VUV spectroscopies \cite{KotaniIdea,KotaniIdea2,WrayNiO,WrayFrontiers,WrayRIXSinterference}. 

\end{thebibliography}
\end{document}